\documentclass[conference,10pt]{IEEEtran}

\usepackage{graphicx} 
\usepackage{subcaption}
\usepackage{amsmath}
\usepackage{mdframed}
\usepackage{xcolor}
\usepackage[linesnumbered,ruled,vlined]{algorithm2e}

\title{Delay-Optimized Admission Control in
5G Using Reinforcement Learning}
\title{Balancing Profit-Delay Trade-Off: A DRL-Based Admission Control for 5G Network Slices}
\title{Prioritizing Latency with Profit: A DRL-Based Admission Control for 5G Network Slices}

\author{Proggya Chakraborty, Aaquib Asrar, Jayasree Sengupta and~Sipra~Das~Bit
}

\DeclareUnicodeCharacter{2212}{-}
\begin{document}

\maketitle

\begin{abstract}
    5G networks enable diverse services such as eMBB, URLLC, and mMTC through network slicing, necessitating intelligent admission control and resource allocation to meet stringent QoS requirements while maximizing Network Service Provider (NSP) profits. However, existing Deep Reinforcement Learning (DRL) frameworks focus primarily on profit optimization without explicitly accounting for service delay, potentially leading to QoS violations for latency-sensitive slices. Moreover, commonly used epsilon-greedy exploration of DRL often results in unstable convergence and suboptimal policy learning. To address these gaps, we propose \textit{DePSAC} -- a \textit{De}lay and \textit{P}rofit-aware \textit{S}lice \textit{A}dmission \textit{C}ontrol scheme. Our DRL-based approach incorporates a delay-aware reward function, where penalties due to service delay incentivize the prioritization of latency-critical slices such as URLLC. Additionally, we employ Boltzmann exploration to achieve smoother and faster convergence. We implement and evaluate \textit{DePSAC} on a simulated 5G core network substrate with realistic Network Slice Request (NSLR) arrival patterns. Experimental results demonstrate that our method outperforms the DSARA baseline in terms of overall profit, reduced URLLC slice delays, improved acceptance rates, and improved resource consumption. These findings validate the effectiveness of the proposed \textit{DePSAC} in achieving better QoS–profit trade-offs for practical 5G network slicing scenarios.

\end{abstract}

\section{Introduction}

The emergence of 5G networks \cite{PB_INDICON'24} has transformed the telecommunication landscape by enabling diverse service categories with heterogeneous requirements, such as enhanced Mobile Broadband (eMBB), Ultra-Reliable Low Latency Communications (URLLC), and massive Machine Type Communications (mMTC). Network slicing \cite{CN'25} has become a critical enabler in 5G to meet these demands, allowing the creation of logically isolated virtual networks over a shared physical infrastructure. However, efficient slice admission control and resource allocation remain key challenges due to dynamic traffic patterns, stringent QoS requirements, and the need for profitability for NSPs. Notably, DRL offers a powerful framework for developing intelligent admission control policies in 5G networks \cite{Hsu_TNSM'25}, balancing long-term utilization while adhering to stringent service-level constraints.


Several studies explore admission control (AC) and resource allocation (RA) in 5G networks. Han et al.~\cite{Bin'20} propose a utility-driven, multi-service based slicing approach using queuing theory, while Raza et al.~\cite{Raza'19} utilize big data analytics for traffic prediction to improve provider profits. However, both approaches lack differentiation in standardized 5G use cases and neglect core network node allocation. Zhang et al.~\cite{Zhang_InfoCom'19} introduce a heuristic VNF placement method to optimize acceptance ratio and throughput, yet without performing joint AC and RA optimization. William et al.~\cite{Villota-JacomeR22} propose reinforcement learning-based SARA and deep reinforcement learning-based DSARA models for joint AC and RA, considering slice acceptance histories to maximize profit. However, both SARA and DSARA do not account for delays in completing slice requests, which is critical for latency sensitive applications such as URLLC. Furthermore, several studies~\cite{Fu_SAC'19, Nassar_IoT'22} investigate the optimization of VNF placement, multi-tenant slice orchestration, and interference-aware embedding for customized services, demonstrating the potential of DRL to manage complex 5G resource allocation tasks \cite{Survey'17}. However, most of the existing work focuses primarily on maximizing profit and/or resource utilization without adequately incorporating QoS metrics such as delay, which are essential for URLLC and other latency-sensitive applications.


This gap motivates us to develop \textit{\textbf{DePSAC}}, a slice admission control scheme that explicitly incorporates service delay while maintaining high NSP profits. Although the baseline DSARA framework~\cite{Villota-JacomeR22} effectively maximizes profit, it does not account for the delays incurred by different types of slices, thereby risking QoS violations, particularly for latency-sensitive services such as URLLC. Furthermore, DSARA relies on an $\epsilon$-greedy exploration strategy, which often results in unstable learning and suboptimal policy convergence due to inefficient state space exploration~\cite{Hao_TNNLS'24}. 

To address these limitations, \textit{\textbf{DePSAC}} introduces a delay-aware reward function to incentivize prioritization of latency-sensitive NSLRs. Furthermore, \textit{\textbf{DePSAC}} leverages Boltzmann exploration, which promotes smoother and more effective convergence by assigning probabilistic action preferences based on learned Q-values.

\noindent The key contributions of this paper are as follows:
\begin{itemize}
    \item We propose a profit and delay-aware reward formulation for DRL-based slice admission control that balances QoS requirements with NSP profitability.
    \item We integrate Boltzmann exploration into the DRL agent to enhance learning stability and prevent local optima convergence issues inherent in $\epsilon$-greedy approaches.
    \item We implement \textit{\textbf{DePSAC}} on a simulated 5G core network substrate with realistic NSLR arrival patterns, and conduct a comparative evaluation against DSARA~\cite{Villota-JacomeR22}.
    \item Experimental results validate that \textit{\textbf{DePSAC}} achieves improved trade-offs between profit and QoS, with shorter service delays, higher acceptance rates, and lower bandwidth utilization over the baseline.
\end{itemize}

The remainder of this paper is organized as follows. Section \ref{sec:prelims} describes the preliminaries and background on 5G network slicing and DSARA. Section \ref{sec:proposal} details our proposed \textit{\textbf{DePSAC}}. Section \ref{sec:res} presents the experimental setup, evaluation metrics, and results. Finally, Section \ref{sec:conclusion} concludes the paper and outlines directions for future work.

\section{Preliminaries and Background \label{sec:prelims}}

\subsection{System Architecture}

We adopt the system architecture from the DeepSARA framework \cite{Villota-JacomeR22}, whose major components are explained here.

\subsubsection{\textbf{5G Core Network Substrate}}

The substrate network is modeled as an NFV infrastructure (NFVI) \cite{Bin'20}, comprising NFVI-PoPs at the core (high capacity) and the edge (low latency) locations. Control plane VNFs are placed on core nodes, whereas latency-sensitive user plane VNFs like URLLC reside on edge nodes. The substrate is represented as a weighted undirected graph $SN = \{N, L\}$ with nodes $N$ (having CPU capacities) and links $L$ (with bandwidth).

\subsubsection{\textbf{5G Network Slice Requests}}



Each NSLR is defined by its slice type (eMBB, URLLC, mMTC), operational time $T_o$, and a graph $G$ of VNFs and virtual links. VNFs carry CPU and placement requirements, while links specify bandwidth demands. NSL graphs follow the control plane (CP) and user plane (UP) separation with AMF, SMF, and UPF components. URLLC slices include backups for high reliability, while mMTC includes more AMFs due to high device density.

\subsubsection{\textbf{System Modules}}

The architecture includes modules:

\begin{itemize}
    \item \textbf{Admission Control Module (ACM):} Uses a DRL agent to assign priority weights to slice types. Based on these weights and arrival times, the prioritizer queues NSLRs in time-windowed batches. The agent selects actions to maximize long-term profit.
    \item \textbf{Resource Allocation Module (RAM):} Maps VNFs to nodes and virtual links to paths based on availability and QoS constraints. The VNFs in the control plane are mapped to core nodes \cite{Raza'19}, while the URLLC user plane VNFs are placed at edge nodes. Backups are placed on separate nodes.  
    \item \textbf{Monitoring Module:} Continuously gathers resource state data and shares it with the ACM and RAM.
    \item \textbf{Lifecycle Module:} Instantiates accepted slices and releases resources upon expiry.
\end{itemize}

\noindent The ACM dequeues NSLRs based on prioritization and interacts with the RAM for resource mapping. Requests are accepted or rejected based on mapping success, and this process continues until the priority queue is empty.




\subsection{Overview on DSARA}

We briefly summarize the baseline DSARA \cite{Villota-JacomeR22}, which is a DRL-based framework for joint admission control and resource allocation in 5G slicing.

\subsubsection{State and Action Space}



A system state $s$ captures the available CPU at the edge/core nodes and the bandwidth across the links. An action $a$ specifies the priority weights (ranging from 0–10) assigned to eMBB, URLLC, and mMTC slices, determining their admission ratio.







\subsubsection{Reward and Learning}

The DRL agent is trained to maximize profit from accepted slices. The reward reflects normalized profit based on slice cost and revenue with respect to full substrate utilization. A DQN-based agent \cite{Zhang_InfoCom'19} is employed, consisting of Evaluation and Target Neural Networks \cite{NN'89} trained on replay memory and periodic updates for stability.




\subsubsection{Exploration and Policy Execution}

The DSARA uses an $\epsilon$-greedy exploration strategy with decaying $\epsilon$, leading to gradually reduced randomness in action selection. It processes requests in time windows, performs prioritization, attempts resource mapping, receives feedback, and updates policy.

\subsubsection{Resource Allocation}

The RAM handles node and link mapping. The nodes are ranked by embedding potential based on the available CPU and link bandwidth. The VNFs are assigned to high embedding potential (EP) nodes and virtual links are embedded along the shortest feasible paths that meet resource and latency constraints.






\section{Proposed DePSAC \label{sec:proposal}}

We now present our proposed model, \textit{\textbf{DePSAC}}—a \textbf{De}lay and \textbf{P}rofit-aware \textbf{S}lice \textbf{A}dmission \textbf{C}ontrol framework that improves over the DSARA baseline \cite{Villota-JacomeR22}. The \textit{\textbf{DePSAC}} introduces a delay-aware reward formulation and adopts Boltzmann exploration to enhance learning stability, addressing the limitations of the DSARA model which overlooks latency sensitivity and uses unstable exploration strategies.



\subsection{Delay-aware Reward Formulation}


The DSARA optimizes admission decisions based solely on monetary profit, without accounting for delays experienced by different slice types. To better handle latency-sensitive requests, such as URLLC, \textit{DePSAC} integrates a penalty term proportional to the service delay and priority of a slice request. This encourages the DRL agent to prioritize slices with stricter QoS requirements in terms of user satisfaction by balancing delay reduction with profit optimization. For the $i^{th}$ request, net reward ($R$) is calculated as:
\begin{subequations} \label{eq:system}    
    \begin{align}
        {penalty}_i = {priority}_i * {delay}_i \label{eq:penalty}\\
        {profit}_i = ({revenue}_i - {cost}_i) * T_o \label{eq:profit} \\
        reward({nsl}_i) = {profit}_i - {penalty}_i \label{eq:reward}\\
        R = \frac{\sum_{i=0}^{k}reward({nsl}_i)}{maxProfit (SN,T)} \label{eq:R}
    \end{align}
\end{subequations}
\noindent Here, $priority_i$ is determined based on the type of the slice (e.g., URLLC $>$ eMBB $>$ mMTC), and $delay_i$ denotes the time elapsed between the arrival and service of the NSL request $i$. The term $T_o$ represents the operational time of the slice, ${revenue}_i$ refers to the revenue earned by the NSP to accept and service a request, while ${cost}_i$ represents its operational cost on the substrate network $SN$. Lastly, $maxProfit(SN,T)$ denotes the maximum profit achievable when all resources in $SN$ are fully utilized over a period of time $T$. This reward formulation ensures that latency-critical slices are rewarded more when served quickly, guiding the agent toward QoS-aware admission strategies.

\subsection{Boltzmann Exploration for Stability}

DePSAC employs Boltzmann exploration, where probability of action selection depends exponentially on the Q-values \cite{Hao_TNNLS'24}:
\begin{equation}
    P(a) = \frac{e^{Q[s,a]/\tau} \times Q[s,a]}{\sum_a e^{Q[s,a]/\tau}} \label{eq:Pa}
\end{equation}
\noindent where, $Q[s,a]$ denotes Q-value for the action $a$ in state $s$. The temperature parameter $\tau$ regulates the diversity of exploration: higher $\tau$ encourages exploration, while lower values promote exploitation. This method offers smoother convergence by prioritizing promising actions without purely greedy behavior, improving learning stability.


\subsubsection{Slice Admission Control Workflow}

The controller initializes a simulation environment with a trained DRL agent. During each run, NSLRs arrive dynamically and are added to a request window, simulating stochastic real-world traffic. The ACM evaluates each request based on its resource demands and delay sensitivity. Requests are prioritized according to a learned weight vector corresponding to slice types (eMBB, URLLC, mMTC), and decisions are made to accept or reject each slice.

Upon acceptance, the required CPU and bandwidth resources are allocated by the RAM. A reward is computed for the agent based on Eq.~(\ref{eq:reward}), capturing both profit and delay penalties. The agent stores experiences in replay memory and periodically updates its Evaluation Network to refine its policy. Once a slice completes its operational time, the allocated resources are de-allocated and reused for future requests. This loop of prioritization, admission, reward feedback, and policy refinement continues, allowing the agent to learn effective policies that balance towards reducing service delay and increasing NSP profitability.

\subsection{Algorithm Description}

The complete interaction between the modules is described below, and the procedural steps are outlined in Algorithm~\ref{Algo}.


\begin{algorithm}[!tbh]
\DontPrintSemicolon
\SetAlgoLined
\SetKwInOut{Input}{Input}
\SetKwInOut{Output}{Output}
\caption{Proposed \textit{DePSAC} Algorithm}
\label{Algo}
\Input{NSLR ${nsl}_i$, substrate network $SN$}
\Output{Net normalized reward $R$ for a $nsl_i$}
Compute $cost_i \leftarrow$ operational cost of servicing $nsli$\;
Compute $revenue_i \leftarrow$ revenue from servicing $nsli$\;
Compute $delay_i \leftarrow (execution - incoming)_{{nsl}_i}$ time\;
\tcc{Boltzmann Exploration Step}
        Select action $a_t = \{p_{embb}, p_{urllc}, p_{mmtc}\}$ using Eq. (\ref{eq:Pa})\;
        Assign prioritization weight $p_{stype}$ based on $a_t$\;
Set priority of an ${nsl}_i$ request as:\;
\If{($stype$ == URLLC)}{
    $priority_i \leftarrow highest \times p_{urllc}$\;
}
\ElseIf{($stype$ == eMBB)}{
    $priority_i \leftarrow medium \times p_{embb}$\;
}
\Else{
    $priority_i \leftarrow lowest \times p_{mmtc}$\;
}
Compute ${penalty}_i$ for delay using Eq.(\ref{eq:penalty})\; 
Compute agent $profit_i$ using Eq.(\ref{eq:profit})\;
Compute agent $reward_i$ for servicing ${nsl}_i$ via Eq.(\ref{eq:reward})\;
Compute net normalized reward $R$ using Eq.(\ref{eq:R})\;
\Return $R$\;

\end{algorithm}

\noindent \textbf{Controller:} Instantiates the ACM and initializes a simulation environment. A new DRL agent is also created to learn optimal admission strategies over time.

\noindent \textbf{Simulator:} Dynamically generates NSLRs and adds them to a request window ($nsl_{win}$) to emulate real-time traffic patterns. Each NSLR includes type, resource demands, and operational time.

\noindent \textbf{\textit{Event\_Prioritizer()} Function:} It sorts incoming NSLRs into a prioritized list $nsl_{win}$ according to their slice type and arrival time. Prioritization folkows delay sensitivity, with URLLC slices ranked higher than eMBB and mMTC.

\noindent \textbf{\textit{Resource\_Allocator()}:} Allocates CPU and bandwidth to an NLSR, if sufficient resources are available. CPU nodes are ranked by their available processing capacity and degree of connectivity. Incoming requests are then assigned to substrate nodes and links based on their priority in the window. The accepted NSLRs are mapped accordingly and the substrate state is updated:
\begin{align*}
    {SN}_{cpu} & = {SN}_{cpu} - {cpu}_i \\
    {SN}_{bw}  & = {SN}_{bw} - {bw}_i
\end{align*}
\noindent where, $SN_{cpu}$ and $SN_{bw}$ denote the available processing capacity and bandwidth of SN before acceptance, while $cpu_i$ and $bw_i$ represent the processing and bandwidth requirements of request ${nsl}_i$.

\noindent \textbf{\textit{Calculate\_Reward()} Function:} It calculate the agent reward using Eq.\ref{eq:reward}. Lower delays yield higher rewards, incentivizing the agent to prioritize latency-sensitive requests. The agent transitions between states in the simulation environment based on the net reward ($R$) obtained after processing requests in ${nsl}_{win}$ (Eq. \ref{eq:R}).

\noindent \textbf{\textit{Deep\_Q\_Learning()} Function:} It allows the agent to select actions based on observed states and received rewards. For exploitation, the agent selects the action with the highest Q-value in the current state. For exploration, it samples an action using the Boltzmann policy defined in Eq. \ref{eq:Pa}. Formally,
\[
    a= 
\begin{cases}
    {max}_a Q_t(s_t,a),& \text{if } r_n > \epsilon\\
    P(a),              & \text{otherwise}
\end{cases}
\]
\noindent where $r_n$ is a random sample and $\epsilon$ is the exploration threshold. 

\noindent \textbf{\textit{Resource\_Deallocation()} Function:} It releases CPU and bandwidth assigned to completed NSLRs after their operational period ends, restoring substrate availability.



\section{Performance Evaluation} \label{sec:res}

This section presents the simulation environment and evaluates the performance of the proposed \textit{\textbf{DePSAC}} scheme. 




\subsection{Simulation Environment}

The proposed \textit{\textbf{DePSAC}} framework is implemented in Python 3 using a modular, object-oriented design. All modules follow the implementation details outlined in Section~\ref{sec:proposal}. The experiments are carried out on a desktop system running Windows 11, equipped with an AMD Ryzen 5 processor and 16~GB of RAM. The simulation employs the NetworkX library to generate and manage graph-based representations of both the substrate network and incoming NSLRs. For consistency with the baseline DSARA evaluation~\cite{Villota-JacomeR22}, the substrate is modeled as a 64-node Barabási–Albert topology, capturing the scale-free characteristics commonly observed in real-world 5G infrastructures.

The experimental evaluation uses a discrete event simulator that integrates the proposed delay-aware DRL agent with Boltzmann exploration. This environment enables systematic testing under varying traffic intensities, slice compositions, and resource constraints, allowing for reproducible and comparative evaluation of admission control strategies.


\subsection{Simulation Metrics}

The performance of the proposed \textit{\textbf{DePSAC}} framework is evaluated using the following key metrics:
\begin{itemize}
    \item \textbf{Profit:} Profit is defined as the total revenue earned by the NSP by servicing a network slice request, minus the operational cost incurred for allocating processing and bandwidth resources as shown in Eq. (\ref{eq:profit}). 
    \item \textbf{Acceptance Rate (AR):} It represents the proportion of incoming NSLRs that are successfully admitted. Thus, $AR = \frac{{req}_a}{{req}_t}$, where ${req}_a$ is the number of admitted NSLRs and ${req}_t$ is the total number of requests. 
    \item \textbf{Delay:} It measures the time taken to service a request after its arrival. Alternatively, for a request, $Delay = (T_{finished} - T_{arrival})$, where $T_{arrival}$ is the request arrival time and $T_{finished}$ is the time at which servicing is completed. 
    \item \textbf{Resource Consumption:} It refers to the proportion of processing and bandwidth resources available to the substrate network that are actually allocated to serve accepted NSLRs over a given time period.
    \begin{align*}
        C = \frac{\frac{\sum_i bw({nsl}_i)}{BW(SN)}+\frac{\sum_i cpu({nsl}_i)}{CPU(SN)}}{2}
    \end{align*}
    where $BW(SN)$ and $CPU(SN)$ are the total bandwidth and processing capacity of the substrate network, while $\sum_i bw(nsl_i)$ and $\sum_i cpu(nsl_i)$ represent the total bandwidth and CPU consumed by admitted slices.

    
\end{itemize}
These metrics collectively assess the effectiveness of the proposed \textit{DePSAC} in balancing economic profit, QoS compliance, admission efficiency, and resource minimization.

\subsection{Results and Analysis}

This section presents the results of our experimental evaluation, comparing the proposed \textit{\textbf{DePSAC}} framework against the baseline DSARA model \cite{Villota-JacomeR22}. To comprehensively assess the impact of our enhancements, we have conducted four set of experiments focusing on key performance metrics. We report both quantitative improvements and qualitative observations to highlight the operational behavior of the proposed \textbf{\textit{DePSAC}}.



\begin{figure*}[!t]
\begin{minipage}[b]{0.32\linewidth}
\centering
\fbox{\includegraphics[width=0.89\textwidth]{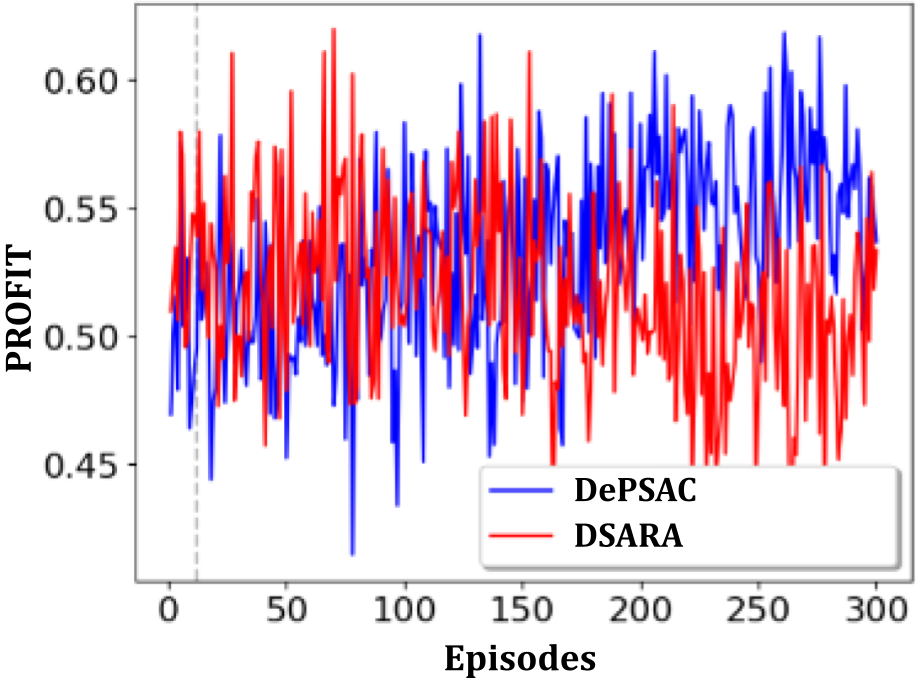}}
\caption{\small \sl Overall Profit}
\label{fig:profit}
\end{minipage}
\hspace{0.1cm}
\begin{minipage}[b]{0.32\linewidth}
\centering
\fbox{\includegraphics[width=0.97\textwidth]{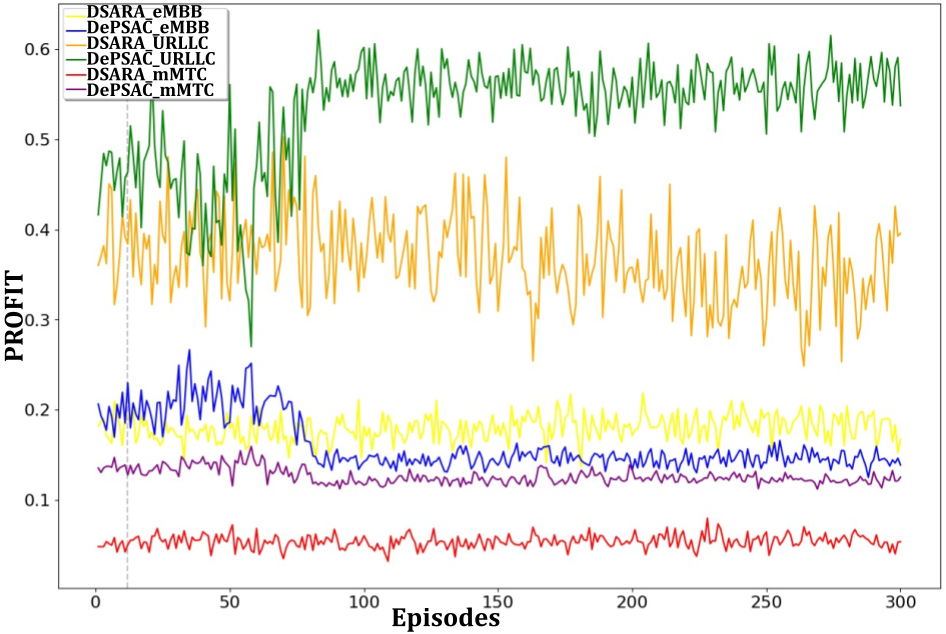}}
\caption{\small \sl Profit for each use-case}
\label{fig:NSLR_profit}
\end{minipage}
\hspace{0.1cm}
\begin{minipage}[b]{0.32\linewidth}
\centering
\fbox{\includegraphics[width=0.98\textwidth]{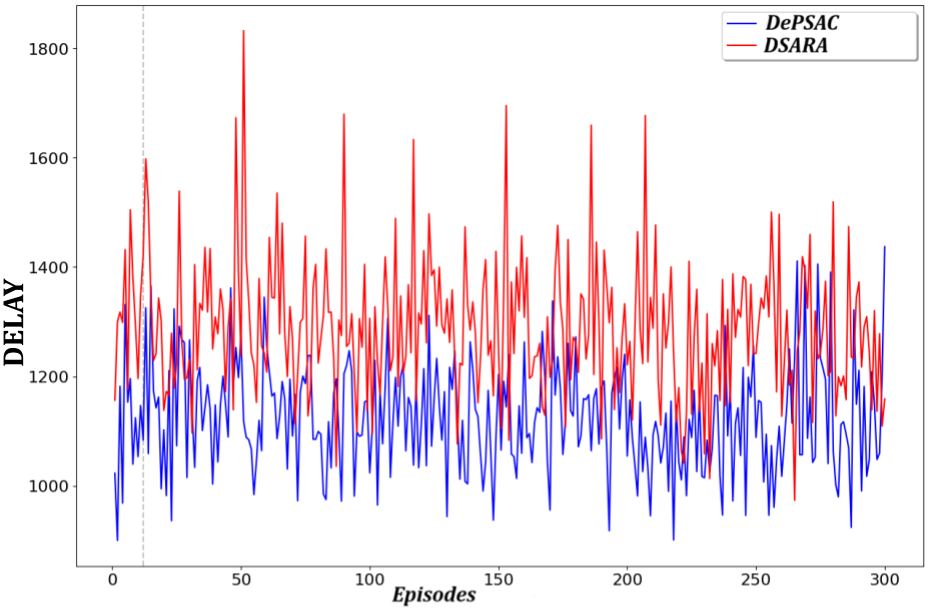}}
\caption{\small \sl Overall Delay} 
\label{fig:delay}
\end{minipage}
\end{figure*}

\noindent\textbf{Profit:} The first set of experiments evaluates the overall profit achieved by the proposed \textit{\textbf{DePSAC}} model over time, as shown in Fig.~\ref{fig:profit}. In the initial episodes, \textit{DePSAC} yields slightly lower profits compared to the baseline DSARA model \cite{Villota-JacomeR22} due to the agent's ongoing exploration under the delay-penalized reward formulation. This behavior is expected, as the agent learns to adapt to the new trade-offs between profit and delay. As training progresses, \textit{\textbf{DePSAC}} consistently outperforms the baseline in terms of overall profit. These results demonstrate that the introduction of a delay penalty does not compromise profitability. Instead, the agent learns policies that effectively balance or even improve NSP revenue maximization.

Furthermore, Fig.~\ref{fig:NSLR_profit} presents the profit evolution for each service type—eMBB, URLLC, and mMTC over time. The results indicate that \textit{\textbf{DePSAC}} consistently achieves higher profits for all service categories. Notably, URLLC slices experience the most significant gain, highlighting the model’s ability to prioritize latency-sensitive, services. Meanwhile, profits from eMBB and mMTC slices are also maintained or improved, indicating a well-balanced allocation strategy. 

\begin{figure*}[!t]
    \centering
    \begin{subfigure}[t]{0.33\textwidth}
        \centering
        \fbox{\includegraphics[width = 0.95\linewidth]{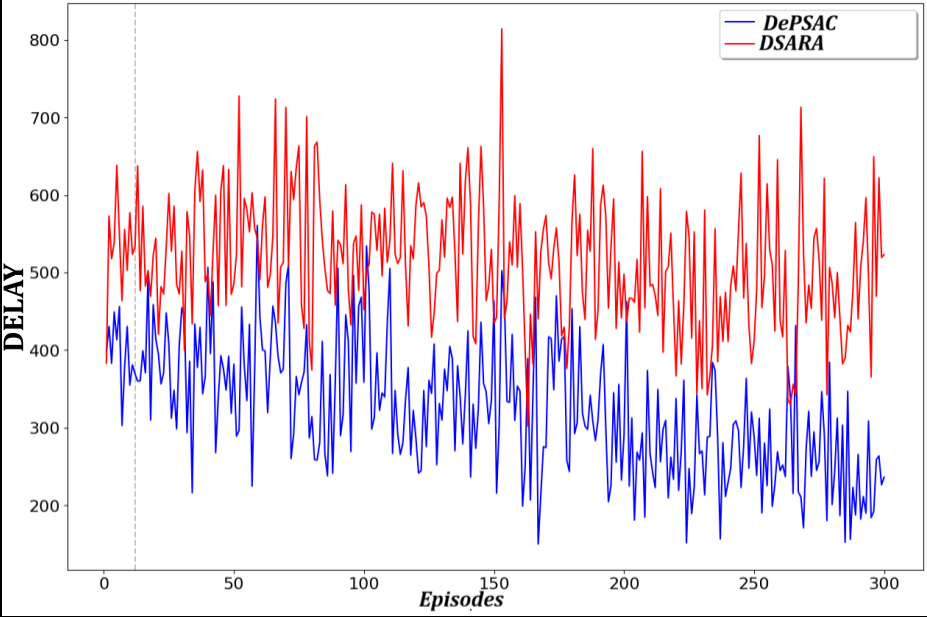}}
        \caption{\small Delay for URLLC}
        \label{fig:urrlc_delay}
    \end{subfigure}%
    \begin{subfigure}[t]{0.33\textwidth}
        \centering
     \fbox{\includegraphics[width = 0.94\linewidth]{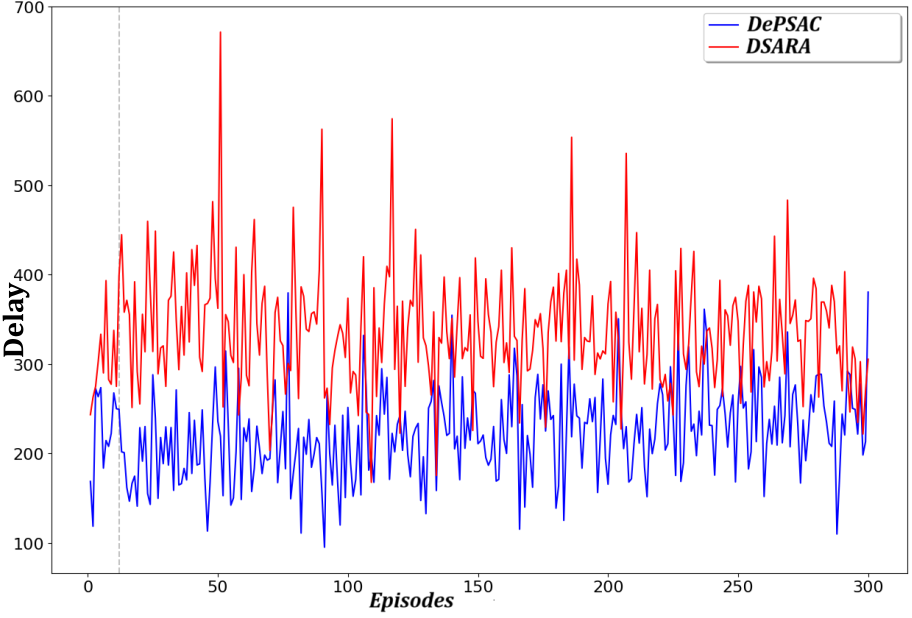}}
        \caption{\small Delay for mMTC}
        \label{fig:miot_delay}
    \end{subfigure}
    \begin{subfigure}[t]{0.33\textwidth}
        \centering
     \fbox{\includegraphics[width = 0.95\linewidth]{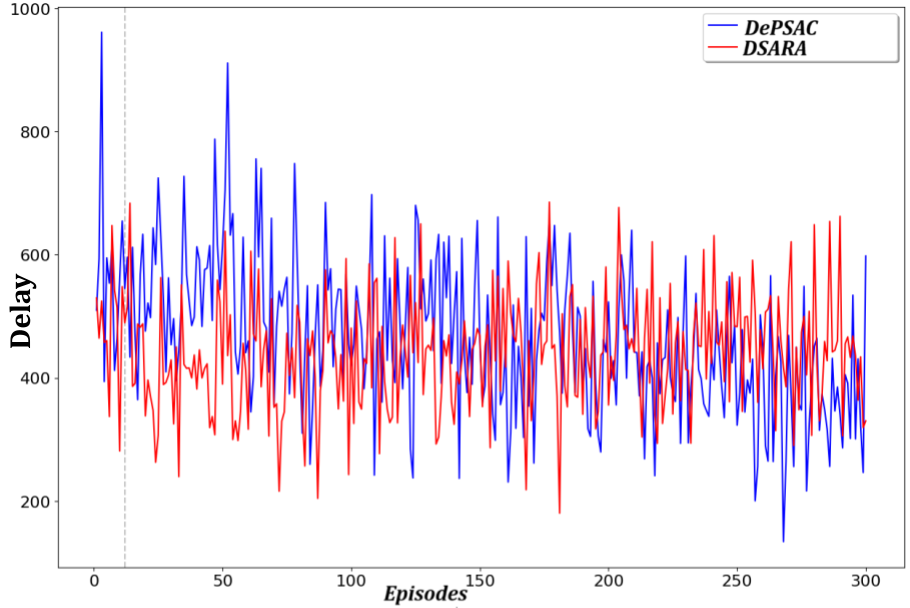}}
        \caption{\small Delay for eMBB}
        \label{fig:embb_Delay}
    \end{subfigure}
    \caption{\small Delay across each use-case} \label{fig:delays}
    \vspace{-1.5em}
\end{figure*}

\noindent\textbf{Delay:} In the next set of experiments, we evaluate (Fig.~\ref{fig:delay}) the overall delay experienced across all service types. The results reveal that \textit{\textbf{DePSAC}} consistently achieves a lower average delay compared to  DSARA. This improvement comes from the delay-aware reward formulation, which penalizes prolonged service times and thereby encourages the agent to prioritize faster request handling. Consequently, the agent learns to make admission and allocation decisions that reduce end-to-end delay, particularly by benefiting latency-sensitive slices. 

To further analyze delay behavior by service type, Fig.~\ref{fig:delays} plots individual delay trends over time. For URLLC (Fig.~\ref{fig:urrlc_delay}), \textit{\textbf{DePSAC}} achieves a substantial reduction in delay relative to DSARA, confirming its effective prioritization of time-critical slices. For mMTC (Fig.~\ref{fig:miot_delay}), delay reductions are moderate yet consistent, indicating improved scheduling without sacrificing lower priority requests. Lastly, for eMBB (Fig.~\ref{fig:embb_Delay}), the delay initially increases slightly due to exploration but converges to values lower than baseline as training progresses. These results demonstrate that \textit{\textbf{DePSAC}} successfully minimizes delays for high-priority slices while balancing delay reduction across other service types under constrained resources, resulting in improved QoS compliance system-wide.

\begin{figure*}[!t]
\begin{minipage}[b]{0.32\linewidth}
\centering
\fbox{\includegraphics[width=0.98\textwidth]{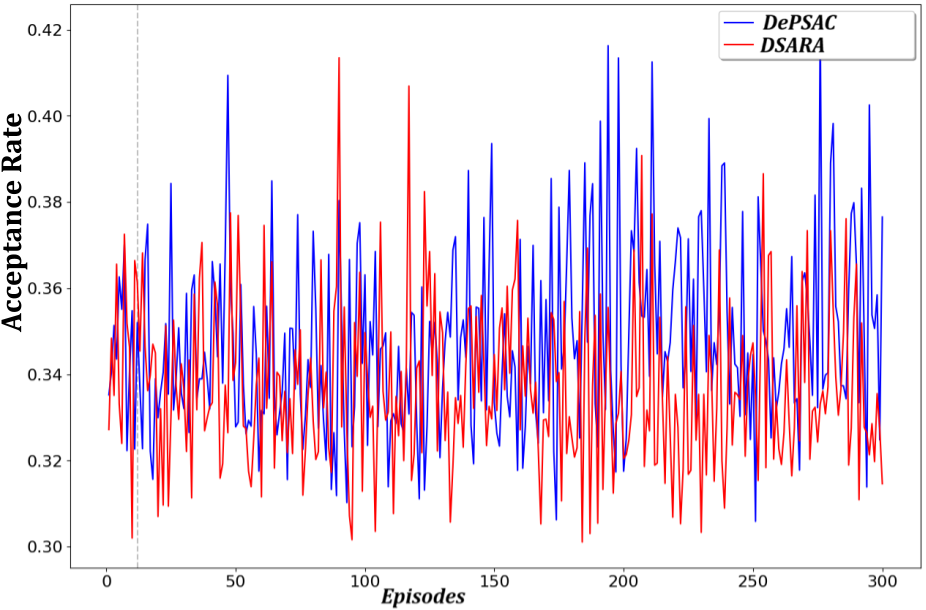}}
\caption{\small \sl Overall Acceptance Rate} \label{fig:acceptance}
\end{minipage}
\hspace{0.1cm}
\begin{minipage}[b]{0.32\linewidth}
\centering
\fbox{\includegraphics[width=0.97\textwidth]{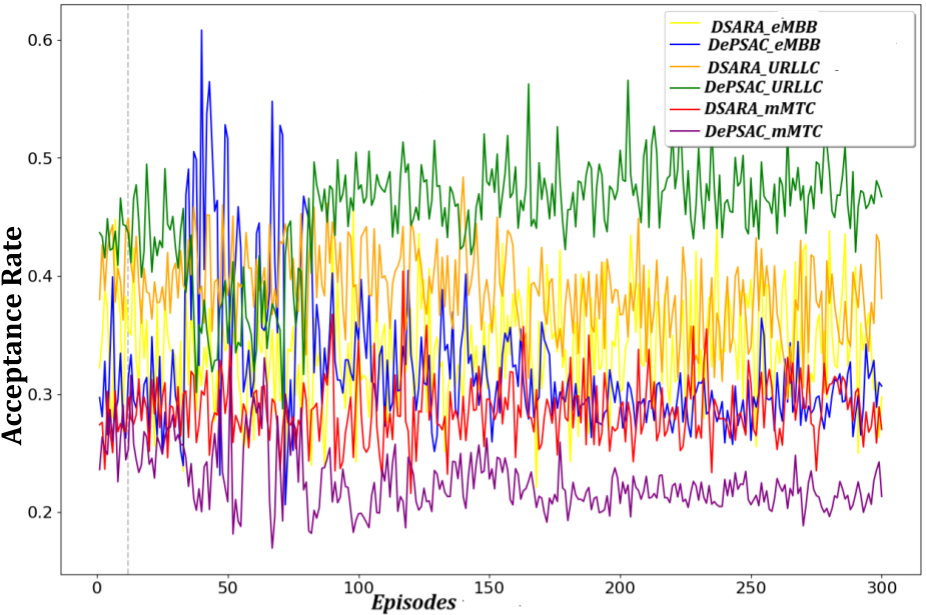}}
\caption{\small \sl Acceptance Rate for each use-case} \label{fig:acceptance_use-case}
\end{minipage}
\hspace{0.1cm}
\begin{minipage}[b]{0.32\linewidth}
\centering
\fbox{\includegraphics[width=0.98\textwidth]{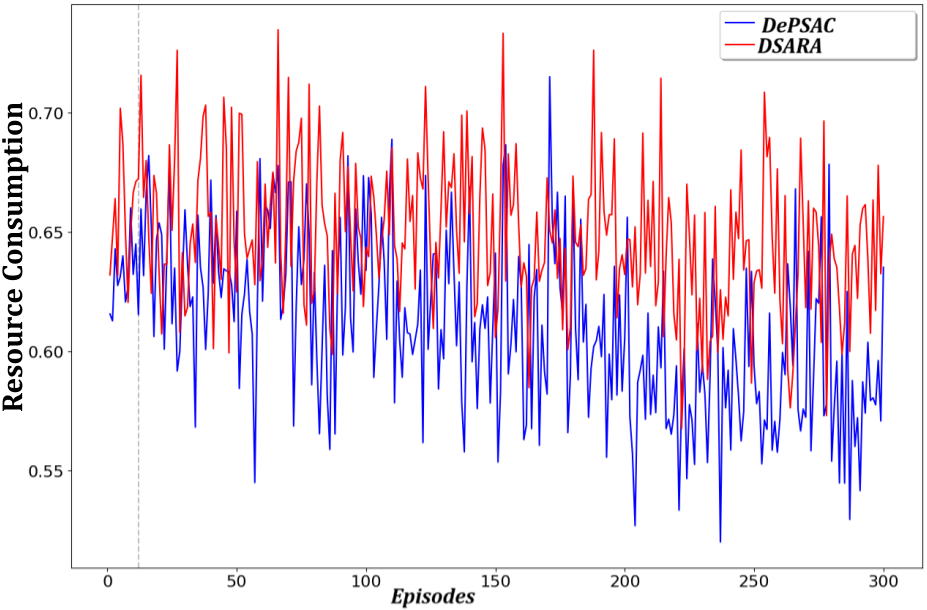}}
\caption{\small \sl Overall Resource Consumption} \label{fig:resource_utilization}
\end{minipage}
\end{figure*}

\noindent\textbf{Acceptance Rate:} In the following set of experiments, we evaluate the overall acceptance rate of NSLRs over time, as shown in Fig.~\ref{fig:acceptance}. Initially, the acceptance rate of \textit{\textbf{DePSAC}} catches up closely to that of DSARA. However, as training progresses, \textit{\textbf{DePSAC}} exhibits a marked improvement, ultimately surpassing the baseline. This gain comes from the delay-aware prioritization mechanism, which enables faster request servicing and, in turn, frees up resources more efficiently, allowing more incoming requests to be admitted over time. 

To understand the slice-specific behavior, we analyze acceptance rates by service type in Fig.~\ref{fig:acceptance_use-case}. For URLLC, \textit{\textbf{DePSAC}} shows a significant improvement over DSARA, reflecting the agent’s learned prioritization of latency-sensitive requests. The eMBB also benefits, with moderately higher acceptance rates, albeit to a lesser extent than URLLC, indicating that the agent maintains strong admission performance for high-throughput slices. Although mMTC acceptance sees a slight decline compared to the baseline, the trade-off remains minimal and within operational thresholds. Collectively, these results confirm that \textit{\textbf{DePSAC}} effectively balances admission control across service types, prioritizing delay-sensitive slices without compromising overall system efficiency.

\begin{figure} [!ht]
\begin{center}  
\fbox{\includegraphics[width=0.70\columnwidth]{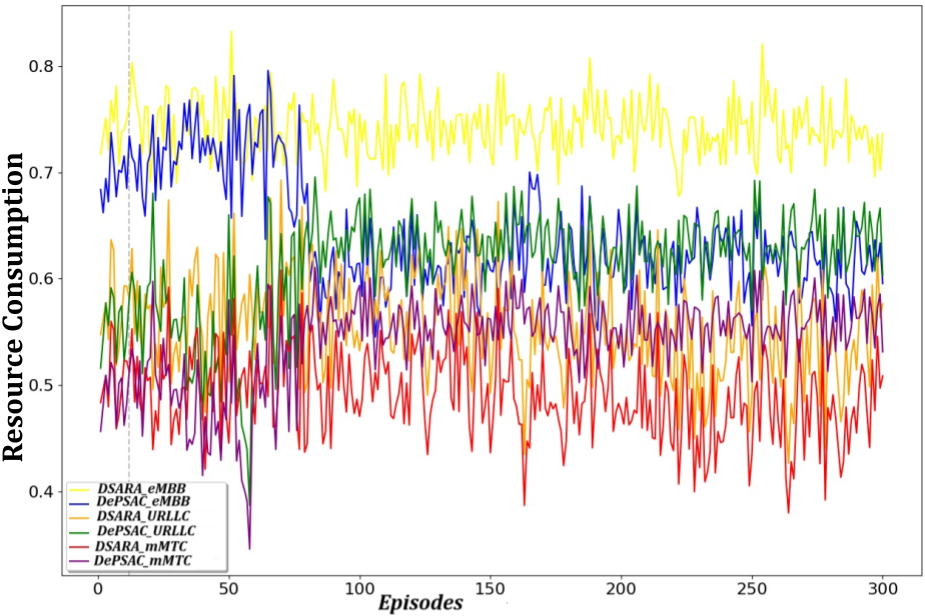}}
\caption{\small \sl Resource Consumption for each use-case \label{fig:resource_utilization_use-case}}
\end{center}  
\end{figure}


\noindent\textbf{Resource Consumption:} In the next set of experiments, we evaluate the overall resource consumption of the substrate network over time, as shown in Fig.~\ref{fig:resource_utilization}. Initially, both \textit{\textbf{DePSAC}} and DSARA exhibit similar levels of CPU and bandwidth consumption. However, in later episodes, \textit{\textbf{DePSAC}} demonstrates a slight reduction in overall consumption. 
The agent increasingly prioritizes URLLC slices, which are less resource-demanding and are generally allocated to edge nodes. This prioritization leads lower aggregate bandwidth usage, an advantageous side-effect of delay-aware decision-making. Furthermore, this reduction is also driven by the lower acceptance rate of eMBB requests, which are highly bandwidth-intensive and demand significant resources from both core and edge nodes. While CPU consumption remains consistent or even improves slightly, the reduced bandwidth footprint indicates more compact and latency-efficient embeddings. 

We further analyze resource consumption disaggregated by slice type and the results are shown in Fig.~\ref{fig:resource_utilization_use-case}. For eMBB, resource usage decreases in line with its reduced acceptance rate, as seen earlier in Fig.~\ref{fig:acceptance_use-case}. In contrast, for URLLC resource consumption increases substantially, demonstrating the agent's effective prioritization of latency-critical services. Interestingly, although URLLC slice admissions increase, total bandwidth utilization remains low due to their lower computational demands. The mMTC slices exhibit stable behavior, with marginal increases in CPU and bandwidth usage, suggesting that admission decisions remain balanced across lower-priority slices as well. Overall, these trends confirm that \textit{\textbf{DePSAC}} achieves more intelligent resource consumption by accelerating the admission of lightweight, delay-sensitive URLLC slices and reducing bandwidth consumption, while maintaining strong CPU utilization and QoS performance.

\begin{mdframed}[backgroundcolor=gray!20, linewidth=0.7pt]

\textbf{Takeaway:} \textit{The experimental evaluation confirms that \textbf{DePSAC}'s delay-aware reward design enables the DRL agent to effectively balance profitability and QoS objectives. The integration of delay penalties incentivizes faster servicing of latency-sensitive URLLC slices without sacrificing overall NSP revenue. Boltzmann exploration fosters smoother and more stable convergence, improving policy learning compared to $epsilon$-greedy strategies. Higher acceptance rates and increased URLLC resource consumption, along with reduced bandwidth consumption, lead to intelligent decision making. The simulation results validate that \textbf{DePSAC} not only improves economic efficiency but also aligns well with 5G's low-latency service requirements. The proposed framework holds promise for generalization to broader multi-objective network optimization scenarios.}

\end{mdframed}



\section{Conclusion \label{sec:conclusion}}
In this paper, we present \textit{\textbf{DePSAC}} that is aware of delay and profit for slice admission control in 5G networks using deep reinforcement learning. By incorporating a delay-penalized reward function and replacing $epsilon$-greedy with Boltzmann exploration, \textit{\textbf{DePSAC}} effectively addresses the QoS limitations of the baseline DSARA model. Our approach enables the agent to prioritize latency-critical slices, such as URLLC, without compromising overall NSP profitability. Comprehensive experiments on a simulated 5G core network show that \textit{\textbf{DePSAC}} consistently outperforms DSARA across multiple performance metrics, achieving higher profits, reduced delays, improved acceptance rates and more efficient resource utilization. These results highlight the importance of delay-aware policy learning in ensuring service differentiation and QoS compliance in network slicing environments. As future work, our aim is to extend \textit{\textbf{DePSAC}} to support federated 5G architectures and evaluate its robustness under dynamic traffic loads, mobility using real-world deployment traces.

\bibliographystyle{unsrt}
{\footnotesize
\bibliography{references}}

\begin{thebibliography}{10}

\bibitem{PB_INDICON'24}
Varsha Poddar and Sipra~Das Bit.
\newblock {QoS Aware Energy Efficient Call Admission Control for 5G Networks}.
\newblock In {\em 2024 IEEE 21st India Council International Conference (INDICON)}, pages 1--6, 2024.

\bibitem{CN'25}
Vidhya P, Subashini K, Sathishkannan R, and Gayathri S.
\newblock {Dynamic network slicing based resource management and service aware Virtual Network Function {(VNF)} migration in 5G networks}.
\newblock {\em Computer Networks}, 259, 2025.

\bibitem{Hsu_TNSM'25}
Yi-Huai Hsu, Chen-Fan Chang, and Chao-Hung Lee.
\newblock {A DRL Based Spectrum Sharing Scheme for multi-MNO in 5G and Beyond}.
\newblock {\em IEEE Transactions on Network and Service Management}, 2025.

\bibitem{Bin'20}
Bin Han, Vincenzo Sciancalepore, Xavier Costa-Pérez, Di~Feng, and Hans~D. Schotten.
\newblock {Multiservice-Based Network Slicing Orchestration With Impatient Tenants}.
\newblock {\em IEEE Transactions on Wireless Communications}, 19(7):5010--5024, 2020.

\bibitem{Raza'19}
Muhammad~Rehan Raza, Ahmad Rostami, Lena Wosinska, and Paolo Monti.
\newblock {A Slice Admission Policy Based on Big Data Analytics for Multi-Tenant 5G Networks}.
\newblock {\em Journal of Lightwave Technology}, 37(7):1690--1697, 2019.

\bibitem{Zhang_InfoCom'19}
Qixia Zhang, Fangming Liu, and Chaobing Zeng.
\newblock {Adaptive Interference-Aware VNF Placement for Service-Customized 5G Network Slices}.
\newblock In {\em IEEE INFOCOM}, pages 2449--2457, 2019.

\bibitem{Villota-JacomeR22}
William~Fernando Villota{-}Jacome, Oscar Maur{\'{\i}}cio~Caicedo Rend{\'{o}}n, and Nelson L.~S. da~Fonseca.
\newblock {Admission Control for 5G Core Network Slicing Based on Deep Reinforcement Learning}.
\newblock {\em {IEEE} Systems Journal}, 16(3):4686--4697, 2022.

\bibitem{Fu_SAC'19}
Fu~Xiao, Lei Chen, Hai Zhu, Richang Hong, and Ruchuan Wang.
\newblock {Anomaly-Tolerant Network Traffic Estimation via Noise-Immune Temporal Matrix Completion Model}.
\newblock {\em IEEE Journal on Selected Areas in Communications}, 37(6):1192--1204, 2019.

\bibitem{Nassar_IoT'22}
Almuthanna~T. Nassar and Yasin Yilmaz.
\newblock {Deep Reinforcement Learning for Adaptive Network Slicing in 5G for Intelligent Vehicular Systems and Smart Cities}.
\newblock {\em {IEEE} Internet Things J.}, 9(1):222--235, 2022.

\bibitem{Survey'17}
Kai Arulkumaran, Marc~Peter Deisenroth, Miles Brundage, and Anil~Anthony Bharath.
\newblock {Deep Reinforcement Learning: A Brief Survey}.
\newblock {\em IEEE Signal Processing Magazine}, 34(6):26--38, 2017.

\bibitem{Hao_TNNLS'24}
Jianye Hao, Tianpei Yang, Hongyao Tang, Chenjia Bai, Jinyi Liu, Zhaopeng Meng, Peng Liu, and Zhen Wang.
\newblock {Exploration in Deep Reinforcement Learning: From Single-Agent to Multiagent Domain}.
\newblock {\em IEEE Transactions on Neural Networks and Learning Systems}, 35(7):8762--8782, 2024.

\bibitem{NN'89}
Kurt Hornik, Maxwell~B. Stinchcombe, and Halbert White.
\newblock {Multilayer feedforward networks are universal approximators}.
\newblock {\em Neural Networks}, 2(5):359--366, 1989.

\end{thebibliography}

\end{document}